\documentclass[a4paper,11pt]{article}
\usepackage[dvipsnames]{xcolor}
\usepackage[utf8]{inputenc}
\usepackage[T1]{fontenc}
\usepackage[bbgreekl]{mathbbol}
\usepackage{geometry}
\usepackage{makecell}

\usepackage{blindtext}
\usepackage{multirow}

\usepackage[natbibapa]{apacite}
\usepackage{xcolor}
\usepackage{amsmath, array, amssymb, amsfonts,amsthm}
\usepackage{upgreek}
\usepackage{xfrac}
\usepackage[inline]{enumitem}
\setlist{nolistsep}
\usepackage[french, english]{babel}
\usepackage[all]{xy}
\usepackage{txfonts}  
\usepackage{sectsty}
\usepackage{booktabs}
\usepackage{caption}
\usepackage{dsfont}
\usepackage{mathtools}
\usepackage{slashed}
\usepackage[makeroom]{cancel}
\usepackage[hidelinks]{hyperref}
\usepackage{textcomp}
\usepackage{multicol}
\usepackage[title]{appendix}
\usepackage{hyperref}
\hypersetup{colorlinks=true, urlcolor=blue, citecolor=blue, linktoc=page}
\allowdisplaybreaks

\hypersetup{
           breaklinks=true,   
           colorlinks=true,   
           pdfusetitle=true,  
        }

\usepackage{tikz}
\usepackage{tikz-cd}
\usetikzlibrary{cd}
\tikzcdset{
arrow style=tikz,
diagrams={>={Straight Barb[scale=0.8]}}
}
\usetikzlibrary{matrix,arrows,decorations.pathmorphing}

\DeclareMathAlphabet{\mathpzc}{OT1}{pzc}{m}{it}

\usepackage{fancybox}
\usepackage{mathrsfs}
\usepackage{scrextend}
\usepackage{mathrsfs}
\usepackage{footmisc}

\usepackage{mathtools}

\makeatletter
\renewcommand*\env@matrix[1][\arraystretch]{%
  \edef\arraystretch{#1}%
  \hskip -\arraycolsep
  \let\@ifnextchar\new@ifnextchar
  \array{*\c@MaxMatrixCols c}}
\makeatother

\newcommand{\defeq}{\vcentcolon=}
\newcommand{\rdefeq}{=\vcentcolon}

\newcommand\M{\mathcal{M}}

\newcommand\RR{\mathbb{R}}
\newcommand\CC{\mathbb{C}}
\newcommand\C{\mathcal{C}}

\newcommand\id{\textit{id}}

\renewcommand\H{\mathcal{H}}

\renewcommand\S{\mathcal{S}}

\newcommand\SU{\mathcal{SU}}
\newcommand\U{\mathcal{U}}
\newcommand\SO{\mathcal{SO}}

\newcommand\K{\mathcal{K}}
\newcommand\J{\mathcal{J}}
\renewcommand\O{\mathcal{O}}
\newcommand\GL{\mathcal{GL}}
\newcommand\D{\mathcal{D}}

\newcommand\vphi{\varphi}

\renewcommand\u{u_{\text{\tiny L}}}

\renewcommand\epsilon{\varepsilon}

\newcommand\rarrow{\rightarrow}

\newcommand\LieG{\mathfrak{g}}
\newcommand\LieH{\mathfrak{h}}

\newcommand\so{\mathfrak{so}}

\newcommand\gl{\mathfrak{gl}}

\newcommand\h{\widehat}
\renewcommand\b{\bar }

\renewcommand\d{\partial}
\newcommand\s{\sigma}
\newcommand\bs{\boldsymbol}

\renewcommand\-{^{-1}}
\newcommand\Ad{\text{Ad}}
\newcommand\ad{\text{ad}}
\renewcommand\id{\text{id}}

\makeatletter

\newcommand{\Rmnum}[1]{\expandafter\@slowromancap\romannumeral #1@}
\makeatother

\makeatletter
\newcommand{\leqnomode}{\tagsleft@true\let\veqno\@@leqno}
\newcommand{\reqnomode}{\tagsleft@false\let\veqno\@@eqno}
\makeatother

\DeclareMathOperator{\Diff}{Diff}
\DeclareMathOperator{\Aut}{Aut}

\newtheorem{thm}{Theorem}

\newtheorem{prop}[thm]{Proposition}
\theoremstyle{definition}
\newtheorem{definition}[thm]{Definition}

\begin{document}

\title{Dressing vs. Fixing: \\ On How to Extract and Interpret Gauge-Invariant Content}
\author{P. Berghofer${\,}^{a}$, J. François${\,}^{a,\,b}$}
\date{}

\maketitle
\begin{center}
\vskip -0.8cm
\noindent
${}^a$ Department of Philosophy -- University of Graz. \\
Heinrichstraße 26/5, 8010 Graz, Austria. \\[2mm]
${}^b$ Department of Mathematics \& Statistics, Masaryk University -- MUNI. \\
Kotlářská 267/2, Veveří, Brno, Czech Republic. \\[2mm]
\end{center}

\vspace{-3mm}

\begin{abstract}
There is solid consensus among physicists and philosophers that, in gauge field theory, 
for a quantity to be physically meaningful or real, it must be gauge-invariant. 
Yet,  every “elementary” field in the Standard Model of particle physics is actually gauge-variant. 
This has led a number of researchers to insist that new manifestly gauge-invariant approaches must be established. 
Indeed, in the foundational literature, dissatisfaction with standard methods for reducing gauge symmetries has been expressed: Spontaneous symmetry breaking is deemed conceptually dubious, while gauge fixing suffers the same limitations and is subject to the same criticisms as  coordinate choices in General Relativity. 

An alternative 
gauge-invariant proposal  was 
recently  introduced in the literature,  the so-called ``dressing field method" (DFM). 
It is a mathematically subtle tool, and unfortunately prone to be confused with simple gauge transformations, hence with standard gauge~fixings.
As a matter of fact,
in the physics literature 
the two
are often conflated, and in the philosophy community some doubts have been raised about whether there is any substantial difference between them.
Clarifying this issue is of special significance 
for anyone interested in both the foundational issues of gauge theories and  their invariant formulation. 
It is thus our objective  to establish as precisely as possible the 
technical and conceptual
distinctions between the DFM and gauge fixing. 
\end{abstract}





\section{Introduction}

Modern physics is written in the language of gauge field theories. 
The Standard Model (SM) of particle physics is a gauge theory in the sense that it rests on internal \textit{local} symmetries. 
General relativity (GR) is a gauge theory in the sense that it rests on an external \textit{local} symmetry. 
Given the pivotal role of gauge (i.e. local) symmetries in modern physics, it is imperative to thoroughly study their conceptual underpinnings (see, e.g., \cite{Berghofer-et-al2023, BradingCastellani2003, Healey2007, Lyre:2004, Rickles2008, Teh2022, Wallace2022a, Wallace2022b, Weatherall2016}). 
One central topic in this context concerns the \textit{ontological} status of gauge symmetries. 
Should they be interpreted as mere mathematical structure of our \textit{descriptions} of reality or do they \textit{represent the structure} of reality? 
Although the answer to this question remains controversial, it has become the received view to say that while symmetries typically relate different physical states (such as in Galileo’s famous ship example), gauge symmetries relate \textit{mathematically distinct} descriptions to the \textit{same physical} state (see \citealp{Berghofer-et-al2023}; \citealp[1233]{Earman2004}; \citealp[3]{Maudlin2002}).

Relatively independently from the position one adopts with respect to the ontology of gauge symmetries, 
there is noticeable consensus among physicists that \textit{physically real quantities must be gauge-invariant}.
James Anderson, for instance, emphasizes that “gauge conditions by themselves have no physical content” \cite[95]{Anderson1967}. 
Jean Zinn-Justin points out that “gauge-dependent” quantities “cannot be associated with physical observables” \cite[546]{ZinnJustin}.
Similarly, Henneaux and Teitelboim conclude that “[p]hysical variables (‘observables’) are then said to be gauge invariant” \cite[3]{HT1992}.   
Accordingly, physicists typically consider gauge symmetries as unphysical redundancies in our description (\citealp[169]{Tong2016} and \citealp[187]{Zee2010}). 
This, however, is a more involved claim. In this paper, we focus on the nearly universally accepted claim that only gauge-invariant quantities can be physically real.

However, if this is so, we find puzzling tensions at the very heart of modern particle physics.
Most notably, the fields and corresponding particles we usually consider to be the building blocks of reality are described  in the Standard Model Lagrangian by ``elementary" fields that are gauge-variant: 
each of them is picked from a \emph{gauge-orbit}, and the physics is indifferent to any particular choice. 
Given that particle-like phenomena are indeed spotted by detectors in various experimental contexts, from eV-scale phenomena
to TeV-scale colliders, with characteristics expected from the gauge theoretical description, we must conclude that the link between the gauge-variant ``Lagrangian" fields and physical fields/particles is quite subtle. 

Where does this leave us? 
To access the physical content of a theory, or its basic variables, there seems to be a need to remove the gauge-variant information, i.e. to ``reduce" gauge symmetries. 
There already exist well-established tools in physics that allow to do this in an  effective way. 
The two most prominent ones are gauge fixing (GF) and spontaneous symmetry breaking (SSB). 

The latter is not meant to address directly the above conceptual problem, but rather to enable a technical workaround to the apparent conflict between the short-range of nuclear interactions, necessitating massive mediating bosons, and  gauge-invariance, which forbids massive gauge fields. 
Despite its well-known empirical success in the electroweak (EW) model, this approach faces philosophical-conceptual problems. 
It has been recognized that it remains mysterious how the breaking of an unphysical gauge symmetry could have a physical impact on our world (see, e.g., \cite{Earman2004}, 1239). Furthermore, it is well-known that Elitzur’s theorem states that local symmetries cannot be spontaneously broken (\cite{Elitzur1975}). This is proven for lattice QFT and expected to be true also in the continuum (see \cite{Maas2019} for an overview).\footnote{Englert himself emphasized in his Nobel lecture that “strictly speaking there is no spontaneous symmetry breaking of a local symmetry” \cite[205]{Englert2014}.}
These concerns circle back to our initial problem, as they seem to call for a gauge-invariant account of the empirical success of the EW model. We will return to this shortly.  

The most widely tool used in connection to our problem 
is  \textit{gauge fixing}.
As we will explain below, it consists in specifying a so-called gauge-fixing condition such that for every gauge orbit there is one representative fulfilling this condition (see \citealp{Boehm2001} and \citealp[58]{Rickles2008}). 
In a guise or another, GF seems pivotal to the quantization of gauge theories. 
Yet, it is an exact analogue to coordinate-choices in GR: it does not remove the gauge-variant information, it merely hides it. 
It is indeed well understood that what is truly physical in GR is coordinate independent; one must check that no prediction depends essentially on a particular coordinate choice. 
Similarly, in Yang-Mills theory what is physical should be independent of the particular choice of gauge.\footnote{Here we understand Yang-Mills theory in the broad sense of a gauge theory based on a Lie group, which includes QED.} 
Hence the necessity to ensure that any  physical result computed is independent of the particular gauge-fixing condition chosen.
A~gauge-invariant formulation would make such prudence unnecessary.\footnote{
We may mention that an  attempt to make sense of how physical d.o.f. are nested within a gauge orbit is the “one true gauge” interpretation -- see e.g. \cite{Healey2001}.  
According to it, within each gauge orbit there would be only one configuration that represents a physically possible state (see \citealp[58]{Rickles2008}). 
The fundamental problem with this view is that it is in principle impossible to empirically determine which  field within a gauge orbit happens to be the one true gauge (see \citealp[367]{maudlin1998healey} and \citealp[49]{Martin2003}).
Accordingly, this interpretation is widely considered “a highly \textit{ad hoc} way of proceeding” \cite[292]{redhead2002gauge}.
The analogy with GR makes the position very unappealing indeed: 
It would be claiming that there is a ``one true" coordinate system in which the physical d.o.f. (fields) are in their true guise. In other words, it amounts to the claim that there are privileged observers in Nature.
A complete negation of the gains made by relativistic physics. 
}


It is thus
the underlying thesis of this paper 
that many conceptual tensions 
such as those raised above
can be avoided by pursuing manifestly gauge-invariant approaches. 
As observed previously, in philosophy of physics, prominent voices have indeed argued for gauge-invariant approaches particularly in the context of the Brout-Englert-Higgs (BEH) mechanism in the EW theory (see  
\cite{Smeenk2006};
\cite{Lyre2008};  \cite{Struyve2011}; and \cite{Friederich2013, Friederich2014}). In physics, while 
it remains the received view that GF, SSB,
and perturbation theory suffice and that manifestly gauge-invariant methods are not necessary, 
the emergence of dissenting approaches such as  
\cite{Frohlich-Morchio-Strocchi80, Frohlich-Morchio-Strocchi81},
\cite{Chernodub2008}, 
\cite{Ilderton-Lavelle-McMullan2010}, 
\cite{Maas2019},
\cite{Sorella-et-al2020}, 
\cite{Sondenheimer2020}
indicate a new trend that takes the requirement of 
manifestly gauge-invariant formulations 
seriously.

One such approach is the ``dressing field method" (DFM).
It has been introduced and developed in  \cite{GaugeInvCompFields}, \cite{Francois2014}, \cite{Francois2021}, \cite{Francois2023-a} -- see also \cite{Gomes-Riello2018, Gomes-et-al2018, Zajac2023, Riello2024}. Some of its philosophical implications are discussed in \cite{Francois2018}, \cite{Teh-et-al2020, Teh-et-al2021}, and in \cite{Berghofer-et-al2023}.
%
%
Broadly speaking, in DFM one identifies a so-called \textit{dressing field} 
whose properties are
such that when the original fields are dressed up with the dressing field, the resulting \textit{dressed fields} are  gauge-invariant variables. 
%
Formally, these dressed variables very much look like gauge-transformed variables. 
This gives rise to a dual issue: 
First, given the formal similarity, the dressing field can be confused for an element of the gauge group of the theory. 
Second, by this confusion, dressed variables may be misconstrued as elements of the gauge-orbit of the undressed fields, hence as their gauge-fixed versions. In other words, one may conflate DFM with standard GF.

Some have suggested this is no conflation, arguing that 
there is no real distinction between gauge fixings and dressings (e.g. Wallace, personal communication). 
It has been further argued that gauge fixing is all we need to extract 
the physical content 
of a theory
(see e.g. \cite{Struyve2011} section 5, \cite{Wallace2024}; \cite{Gomes2024}), which seems to us to be also implicitly assumed by most QFT textbooks.
Were these claims correct, this would  undermine the appeal  of pursuing gauge-invariant approaches to gauge theories.
We disagree with the above position and, given the relevance of the issue to the foundations of gauge theories, we feel it is useful to make the distinction between GF and DFM as precise as possible, both technically and conceptually.
This will make clear where precisely our disagreement lies with the position that tends to conflate dressings and GF, as originally presented by \cite{McMullan-Lavelle97} and whose more or less explicit variants can be found e.g. in the 
the above-mentioned works of Struyve, Wallace, and Gomes.
%

The paper is structured as follows. 
In section \ref{The dressing field method}, 
 we first give a basic account of gauge field theories, as well as a precise definition of gauge fixing. 
This provides the necessary technical background for our discussion of DFM,  which cleanly distinguishes it from GF. 
In section \ref{Gauge fixing vs dressings in the physics literature}, 
we first emphasize how what is usually understood as GF may turn out to be an instance of dressing. We consider the example of the Lorenz gauge. 
Then, we highlight the presence of  dressings in the existing physics literature, stressing a natural interpretation of DFM that is not available for GF, which should in itself deter from conflating the two.

\section{The dressing field method}
\label{The dressing field method}

The dressing field method (DFM) recently emerged with the 
potential of offering a conceptually clear, philosophically motivated, and technically 
clean way of reducing gauge symmetries (see  e.g. \cite{Attard_et_al2017}; \cite{Francois2021}; and  \cite{Berghofer-et-al2023} section 5). 
Symmetry reduction via DFM is achieved by removing gauge-dependence at the level of field variables: gauge-invariant fields are built systematically via a so-called \emph{dressing field}, and the gauge theory rewritten in terms of these invariant variables. 
This procedure, we will argue, is distinct from both SSB and gauge-fixing, and suggests a quite distinctive interpretive picture.


As we will indeed see in section \ref{Gauge fixing vs dressings in the physics literature}, among other examples,
dressing fields have 
been employed not only for gauge-invariant formulations of QED (see \cite{Dirac55}) but also of QCD (see \cite{McMullan-Lavelle97}). The natural interpretation is that, in QED bare electrons are made gauge-invariant by dressing them up with photon fields, while in QCD  quarks and gluons are made gauge-invariant by dressing them up with gluon fields.\footnote{Of course, there is an important disanalogy in that single physical electrons exist while quarks are confined within hadrons. It has been argued that the reason for this lies precisely in the fact that non-perturbative dressings of single quarks are impossible (\citealp{McMullan-Lavelle97}).}



The DFM is most naturally expressed in the geometric language of fiber bundles.  
In~what follows, we will only give a succinct synthesis of its field theoretic version. 
We will limit  
the technical background to the minimum necessary
for conceptual clarity, starting with the basic notions of gauge field theory.
For further technical guidance, see \cite{Berghofer-et-al2023, Francois2021} and the references therein.

\subsection{Basics of classical gauge field theories} 
\label{Basics of classical gauge field theories}

We summarize the basic  structure of a gauge field theory -- for an in-depth  exposition, see e.g. \cite{Hamilton2018}. 
Such a theory, in a $m$-dimensional (region $U$ of) spacetime $M$, based on a Lie group  of symmetry  $H$,  describes the dynamics and interactions of a set of fields $\Phi=\{A, \phi\}$ whose kinematics we now remind. 

\paragraph{Kinematics}
$A$ is a 1-form on $U\subset M$ with values in the Lie algebra $\LieH$  of $H$, which represents the gauge potential. We denote $A \in \Omega^1(U, \LieH)$, and given a coordinate system $\{x^\mu\}$ on $U$, we have the decomposition $A=A_\mu\, dx^{\,\mu}$, and $A_\mu$ is the field object appearing in coordinate formulation of field theory ($A_\mu(x)$ is the value of the field at $x\in U\subset M$). Furthermore, given a basis $\{\tau_a\}$ of $\LieH$, one has $A_\mu =A^a_\mu\, \tau_a$. So, the full index structure of the the gauge potential field is $A^a_\mu$, with $a$ the ``internal" index and $\mu$ the spacetime index.
The field strength of $A$ is the 2-form $F:=dA + \sfrac{1}{2}[A, A] \in \Omega^2(U, \LieH)$: in field-like language, $F^a_{\mu\nu}$.\footnote{For $H=U(1)$, the abelian group of electromagnetism, the bracket term vanishes, so: $F=dA$, i.e. $F_{\mu\nu} =\d_\mu A_\nu - \d_\nu A_\mu$. This is the well-known Maxwell-Faraday tensor describing the electromagnetic (EM) field in terms of the EM 4-vector potential.  }  

The field $\phi$ denotes a field (or a collection of fields) taking values in some  representation space(s) $V$ for $H$: these represent various matter fields (scalar fields among them).
The group morphism $\rho: H \rightarrow GL(V)$, with corresponding Lie algebra morphism $\rho_*: \LieH \rightarrow \mathfrak{gl}(V)$, gives the action of $H$ and $\LieH$ on $V$, thus on $\phi$. 
In practive $V$ is often  a real or complex vector space, supporting an action of the defining matrix representation of  $H$. So we may omit the representation in our notation, writing  terms like $\rho(h)\phi$ and $\rho(A)_* \phi$, simply as $h\phi$ and $A\phi$. 

The interaction between the gauge potential and the matter fields is formalised by the (gauge) covariant derivative, 
$D\phi := d\phi + A\phi$, which implements their \emph{minimal coupling} via the term $A\phi$.

\paragraph{Gauge transformations}
Now, these field variables are subject to the action of the (infinite dimensional) gauge group $\H$ of the theory. It is the set of $H$-valued functions  $\gamma: U \rightarrow H$, $x \mapsto \gamma(x)$ -- 
which then enjoy
pointwise group multiplication $(\gamma \gamma')(x)=\gamma(x) \gamma'(x)$ -- with the  \emph{defining property} that they act on (transform) each other via group conjugation:  any given $\eta \in \H$ is acted upon by any other $\gamma \in \H$ as  $\eta \mapsto \gamma^{-1} \eta\gamma=:\eta^\gamma$.\footnote{The right-hand side of the equality is just a notation defined by the left-hand side, and signifies the action of $\gamma\in \H$ on $\eta$ seen as a ``field" on $U\subset M$.} The gauge group is thus \emph{defined} as
 \begin{align}
 \label{Gauge-group}
\H := \left\{ \gamma, \eta :U \rightarrow H\ |\  \eta^\gamma= \gamma^{-1} \eta\gamma \right\}.
\end{align} 

We seize the occasion to stress a point often too easily overlooked: 
there is no proper mathematical definition of a space of objects without specifying its group of automorphisms. In other words, 
the transformation properties of a mathematical object is an integral part of its definition. 
The mathematics of  gauge theory is no exception. 
The case above illustrates this: When looking at a $H$-valued function, one cannot know whether or not it is an element of $\H$ until its transformation property is ascertained. 

By this same general principle, one must be mindful of the fact that the action of $\H$ on the field variables is part of their very \emph{definition}. This action defines their gauge transformations:
  \begin{align}
  \label{GTgauge-fields}
A\ \mapsto\ A^\gamma:=\gamma^{-1} A \gamma + \gamma^{-1} d \gamma \qquad \text{and} \qquad \phi \ \mapsto \ \phi^\gamma:=\gamma^{-1}\phi. 
\end{align} 
Given the definition of the field strength in terms of $A$, and given the definition of the covariant derivative of $\phi$, \eqref{GTgauge-fields} implies:
 \begin{align}
  \label{GTgauge-fields2}
F\ \mapsto\ F^\gamma=\gamma^{-1} F \gamma \qquad \text{and} \qquad D\phi \mapsto (D\phi)^\gamma :=&\ d\phi^\gamma + A^\gamma \phi^\gamma \\
=& \ \gamma^{-1}D\phi.  \notag
\end{align} 
The covariant derivative of $\phi$ gauge transforms like $\phi$, it preserves its gauge-covariance (hence the name ``\emph{covariant} derivative" for $D$). 
Thus, $\phi$, $D\phi$, and $F$ are gauge-covariant fields -- gauge tensors -- while $A$ is not, given its inhomogeneous transformation property.\footnote{Notice however that, given \eqref{GTgauge-fields},  the difference of two gauge potentials $B:=A -A'$ is a gauge covariant 1-form: it transforms as $B \mapsto B^\gamma=\gamma^{-1} B \gamma$. This shows that the space of gauge potentials is an \emph{affine} space modeled on the vector space of $\mathfrak g$-valued ($\Ad$-) covariant 1-forms.}

The deeper mathematical reason for \eqref{Gauge-group}-\eqref{GTgauge-fields}-\eqref{GTgauge-fields2} is that gauge field theories are based on the differential geometry of connections on fiber bundles.
We  summarize the key notions in Appendix \ref{AppendixA} for the interested reader.
\medskip

Given \eqref{Gauge-group}-\eqref{GTgauge-fields}, the action of the gauge group $\H$ on field space $\Phi$ is a right action: $(A^\eta)^\gamma=A^{\eta\gamma}$ and $(\phi^\eta)^\gamma=\phi^{\eta\gamma}$.
Explicitly, e.g. for $\phi$, this obtains as: $(\phi^\eta)^\gamma=(\eta\- \phi)^\gamma = (\eta^\gamma)\- \phi^\gamma = (\gamma\- \eta \gamma)\- \gamma\- \phi = \gamma\-\eta\- \phi =(\eta\gamma)\-\phi \rdefeq \phi^{\eta\gamma}$.
The $\H$-orbit of $\{A, \phi\}$ is denoted $\O^\H_{\{A, \phi\}}$, or $\O_{\{A, \phi\}}$ when no confusion arises, and is called a gauge orbit. 
The right action of $\H$ foliates $\Phi$ into  gauge orbits $\O$ which, under adequate restrictions on either $\Phi$ (excluding field configurations with stability subgroups) or $\H$ (considering elements s.t.$\,\gamma ={\id_H}_{\,|\d U}$), are isomorphic~to~$\H$. This means that, under these conditions, the field space $\Phi$ is a principal fiber bundle with structure group $\H$ over the moduli (orbits) space $\M\defeq\Phi/\H$: one writes $\Phi \xrightarrow{\pi} \M$, with projection map $\pi(\{A, \phi\})=\{[A], [\phi]\}$. 
For~more on that point, see e.g. \cite{Singer1978, Singer1981, Ashtekar-Lewandowski1994, Baez1994,  Fuchs-et-al1994, Fuchs1995}. 
This~viewpoint is fruitful to understand gauge-fixing, as we shall discuss shortly.

\paragraph{Dynamics}
The Lagrangian of a field theory is  $L:\Phi \rarrow \Omega^m(U)$, $(A, \phi) \mapsto L(A, \phi)$. It is a volume form on $U\subset M$, meant to be integrated to give the action functional $S=\int_U L$, from which one either extracts classical field equations (via the variational principle), or builds a quantum theory. Most often this is done via a path integral $Z(A, \phi)= \int \delta \Phi   \, e^{\sfrac{i}{\hbar}S}$, where $\delta \Phi\defeq \delta A \delta \phi$ is the measure on field space~$\Phi$. 

The gauge principle can be understood as a symmetry principle meant to narrow down the class of admissible theories (and/or select those with desirable properties). A $\H$-gauge theory is one whose Lagrangian $L=L(A, \phi)$ is required to be $\H$-invariant, possibly up-to boundary terms: i.e. $L(A^\gamma, \phi^\gamma)=L(A, \phi) + db(A, \phi; \gamma)$, which one may denote simply $L^\gamma=L+db$. 
This guarantees the $\H$-invariance of the action $S^\gamma=S$,\footnote{If $b\neq 0$, this holds when  $\d U =\emptyset$ or if boundary conditions are imposed: either $\{A,\phi\}=0_{\,|\d U}$ or ${\{\delta A,\delta \phi\}=0}_{\,|\d U}$. 
Preserving these necessitates to  restrict the admissible gauge transformations, e.g. to the subgroup $\H_{\d}\subset \H$ whose elements are s.t. $\gamma={\id_H}_{\,|\d U}$.}  therefore the $\H$-covariance of the field equations. 
The quantum gauge theory is well-defined if  $Z^\gamma=Z$, i.e. if no gauge anomaly arises from the gauge transformation of the measure $\delta\Phi^\gamma$. 

\paragraph{Gauge-fixing}
In classical gauge field theory (mostly classical electrodynamics),  
one might often find it convenient to simplify computations by restricting to those variables $\{A, \phi\}$ satisfying particular functional properties, i.e. some differential equations. 
If such restrictions are imposed by exploiting (even tacitly) the gauge freedom \eqref{GTgauge-fields} of the field variables, they are called ``gauge-fixing." 

In quantum gauge field theory, such restrictions are more than a mere convenience,  they are necessary to even get started:  $Z$ is a priori a divergent quantity, as integration over field space takes into account gauge-related and physically identical configurations, which produces an infinite factor due to the volume of $\H$. 
One~must therefore restrict $Z$ to a ``slice" in $\Phi$, cutting across gauge orbits once, selecting a single representative in each orbit: this is what a gauge-fixing slice is meant to achieve.

A gauge-fixing is thus a choice of local section  of the field space fiber bundle $\Phi$: $\s: \U\subset\M \rarrow \Phi$.
Concretely, it must be specified by a gauge condition taking the form of an algebraic and/or differential equation on the field variables, $\C(A, \phi)=0$. 
Making the use of the gauge freedom \eqref{GTgauge-fields} explicit, it is rather $\C(A^\gamma, \phi^\gamma)=0$. 
The~gauge-fixing slice, i.e. the submanifold 
\begin{align}
\label{GF}
\S\defeq \{(A, \phi) \in \Phi_{|\,\U}\, |\, \C(A^\gamma, \phi^\gamma)=0 \} \subset \Phi
\end{align}
is the image of the local section $\s$. 
In general, there is no such global section (no ``good" gauge-fixing) unless the field space bundle $\Phi$ is trivial, i.e. $\Phi=\M \times \H$.\footnote{
The fact that no global gauge-fixing exists for pure $H=SU(n)$-gauge theories over compact regions of spacetime is known as the Gribov-Singer obstruction (or Gribov ambiguity). See \cite{Singer1978, Singer1981, Fuchs1995}. 
}

\subsection{DFM in a nutshell}
\label{DFM in a nutshell}

We are now ready to define the key notions of the dressing field method. Consider a $\H$-gauge theory with Lagrangian $L(A, \phi)$, based on the Lie group $H$. 

\paragraph{Kinematics}
Suppose there is a subgroup $K \subseteq H$ to which corresponds the gauge subgroup $\K \subset \H$. 
Suppose also there is a group $G$ s.t. either  $H\supseteq G \supseteq K$, or  $G \supseteq H$. 
 
\begin{definition}
\label{Def1}
A $\K$-\emph{dressing field}  is a map $u: M \rarrow G$, i.e. $G$-valued field, \emph{defined} by its $\K$-gauge transformation: 
\begin{align}
\label{GT-dressing}
u^\kappa:=\kappa\- u, \quad \text{ for } \kappa \in \K.
\end{align} 
The space of such $G$-valued $\K$-dressing fields is denoted  $\D r[G, \K]$. 
\end{definition} 
\noindent One may call $\K$ (or $K$) the \emph{equivariance} group of $u$, while $G$ is its \emph{target} group.
Given the existence of a $\K$-dressing field, we have the following

\begin{prop} \normalfont 
\label{Prop1}
Given a gauge potential   $A$  and   gauge-tensorial fields collectively denoted $a=\{\phi, D\phi, F \}$, one may define the following \emph{dressed fields}:
\begin{align}
\label{dressed-fields}
A^u\defeq  u\- A \u + u\- du  \quad \text{ and } \quad a^u\defeq u\-a.
\end{align}  
These are $\K$-invariant, as is easily seen from \eqref{GTgauge-fields} --  replacing $\gamma \rarrow \kappa \in \K$ -- and \eqref{GT-dressing}.
\end{prop}
\noindent Clearly, when $u$ is a $\H$-dressing field, $K=H$, the dressed fields \eqref{dressed-fields} are  $\H$-invariant.

 In particular, the dressed curvature is $F^u=u\- Fu=dA^u+\sfrac{1}{2}[A^u, A^u]$, and satisfies the Bianchi identity $D^{A^u}F^u=0$, where the dressed covariant derivative is $D^{A^u}=d\, + A^u$. 
 A dressed matter field is $\phi^u:=u\-\phi$, its coupling to the dressed potential given by $D^{A^u}\phi^u =u\-D\phi=  d\phi^u + A^u\phi^u$.

 Notice that, 
for the dressings $a^u$ to make sense in the case $G \supset H$,  one must assume that representations of $H$ extend to representations of~$G$.
\smallskip

Let us also stress the fact that a \emph{pure gauge potential} (a.k.a. flat connection) is necessarily given by a $\H$-dressing field: $A_0=udu\-$. By  $u^\gamma =\gamma\- u$, $A_0$ indeed $\H$-transforms as a gauge potential \eqref{GTgauge-fields}. 
A flat gauge potential \emph{is never} 
expressible as $\gamma d \gamma\-$ with $\gamma \in \H$:  given the  transformation properties \eqref{Gauge-group} of gauge group elements, the object $\gamma d \gamma\-$ does not $\H$-transform as a gauge potential. 
This fact should help one recognise a dressing field, hence DFM applications, when pure gauge potentials are used.

\paragraph{Dynamics}
Consider the Lagrangian $L=L(A, \phi)$ of a $\H$-gauge theory, satisfying  quasi-$\H$-invariance as previously specified: $L(A^\gamma, \phi^\gamma)=L(A, \phi) + db(A, \phi; \gamma)$, for $\gamma\in \H$. 
Suppose that one disposes of a $\K$-dressing field $u$, with target group $G\subseteq H$. 
Then, exploiting the quasi-invariance of $L$, which holds as formal algebraic property (unrelated to the transformation property of $\gamma$), we have the following

\begin{prop} \normalfont 
\label{Prop2}
 The dressed Lagrangian is the Lagrangian expressed in terms of the dressed variables \eqref{dressed-fields}:
\begin{equation}
\begin{aligned}
\label{dressed-Lagrangian}
L(A^u, \phi^u)&=L(A, \phi) + db(A, \phi; u), \\[1mm]
\text{write } \ \,   L^u &= L+ db(u). 
\end{aligned} 
\end{equation}
\end{prop}
\noindent From this follows that if $L$ is $\H$-invariant, i.e. $b=0$, 
it can be re-written in terms of the dressed variables: $L(A, \phi)= L(A^u, \phi^u)$, or $L=L^u$. 

\paragraph{Residual gauge symmetry}
Suppose $K$ is a normal subgroup of $H$,\footnote{That is, satisfying $h^{-1}K h=K$ for any $h\in H$.} 
denoted $K \triangleleft H$, then $H/K\rdefeq J$ is a Lie group. 
Correspondingly $\K \triangleleft \H$ and $\J =\H/\K$ is a subgroup of $\H$. 
It follows that the $\K$-invariant dressed fields \eqref{dressed-fields} are $\J$-gauge variables, satisfying \eqref{GTgauge-fields}-\eqref{GTgauge-fields2} under the replacement $\gamma \rarrow \eta \in \J$. Therefore the dressed Lagrangian $L^u=L(A^u, \phi^u)$ is a $\J$-gauge theory. Notice that in  case $L$ is $\H$-invariant so that is can be re-written as $L(A, \phi)= L(A^u, \phi^u)$, this means that it is not actually a $\H$-theory but a $\J$-gauge theory. A fact uncovered once the right field variables \eqref{dressed-fields} are identified/built.

\paragraph{Field-dependent dressing fields}
A key idea of DFM is that a dressing field is not to be introduced by hand, but should rather be identified within, or built from, the existing field variables. 
These are called \emph{field-dependent} dressing fields. 
As~functionals of the original fields, they are understood as maps on field space, 
\begin{equation}
\label{Field-dep-dressing}
\begin{aligned}
u\ \ :\ \  \Phi \ &\rarrow\  \D r[G, \K], \\
    \{A, \phi\} \  &\mapsto\  u=u[A, \phi] .
\end{aligned} 
\end{equation}
The initial field variables $\{A, \phi\}$ encode physical d.o.f. in a redundant way, mixing them with non-physical pure gauge d.o.f.
The dressed variables \eqref{dressed-fields}, now written $\{A^{u[A, \phi]}, \phi^{u[A, \phi]}\}$, are then to be understood as a reshuffling of the d.o.f. of the initial fields that eliminates (part of) the pure gauge d.o.f., getting closer to the true physical d.o.f. 
If $u$ is a $\H$-dressing field, the $\H$-invariant dressed fields faithfully represent the physical  d.o.f. embedded in the initial set of gauge fields.  

Observe furthermore that in general it may be that one (set of) field(s) can be dressed by another,  e.g. $A^{u[\phi]}$ or $\phi^{u[A]}$. 
This fact raises interesting  interpretive consequences, examples of which we will encounter shortly. These consequences are relevant to the main thesis of this paper, which we will now delve into. 

\paragraph{Dressing as distinct from gauge fixing}
Comparing the definitions  \eqref{Gauge-group} of the gauge group and  that \eqref{GT-dressing} of a dressing field, one sees clearly that  $u \notin \K$.  
Therefore,  \eqref{dressed-fields} are \emph{not} gauge transformations, despite the formal resemblance with \eqref{GTgauge-fields}.
This means that, e.g., $A^u$ is no more a gauge potential: for one thing, it doesn't transform like one under $\K$ (being $\K$-invariant). Here we remind the point stressed after eq.\eqref{Gauge-group}.

Notice also that, in case $G\supset H$, the dressed potential is $\LieG$-valued, $A^u \in \Omega(U, \LieG)$, not $\LieH$-valued like $A$, and $F^u \in \Omega^2(U, \LieG)$.\footnote{Even if $\LieH$ is an ideal in $\LieG$, i.e. stable under the action of $G$. This is due to the inhomogeneous term $u^{-1}du$. Notice that, with $\LieH$ an ideal, $F^u \in \Omega^2(U, \LieH)$. }
Such a case notably occurs in the gauge treatment of gravity (without spinors).
Starting from a $\H=\SO(1,3)$-theory with 
$A={A^a}_b={A^a}_{b,\, \mu}dx^{\,\mu} \in \Omega^1\big(U, \so(1,3)\big)$ the spin connection, one finds that the components of the soldering form $e={e^a}_\mu dx^{\,\mu} \in \Omega^1(U,\RR^4)$, the tetrad field ${\bs e}={e^a}_\mu$, is a $GL(4)$-valued Lorentz dressing field. 
Indeed, the soldering form has gauge transformation $e^\gamma =\gamma\- e$,  inherited by the tetrad field: 
in components $({e^a}_\mu)^\gamma ={(\gamma\-)^a}_b {e^b}_\mu$.
\mbox{Considering} the field space  for pure gravity $\Phi=\{A, e\}$, quite trivially one has
\begin{equation}
\label{Field-dep-dressing-gravity}
\begin{aligned}
u\ \ :\ \  \Phi \ &\rarrow\  \D r[GL(4), \SO(1,3)], \\
    \{A, e\} \  &\mapsto\  u[A, e] \defeq{e^a}_\mu = \bs e .
\end{aligned} 
\end{equation}
The dressing of $A$ is 
$A^u =A^{\bs e} = {\bs e}\- A {\bs e} + {\bs e}\- d{\bs e} \in \Omega^1\big(U, \gl(4)\big)$. 
Which is none other than the linear connection 
$\Gamma = {\Gamma^\alpha}_{\beta}= {\Gamma^\alpha}_{\beta,\mu} dx^{\,\mu}$. 
Indeed, the linear connection is not a gauge-fixed spin connection, nor is it its gauge-transformation.\footnote{Only the reverse relation could be thus interpreted. Seing $\Gamma$ as the gauge potential for the $\GL(4)$-gauge group, the reduction to $A$ in a $\SO(1,3)$-theory is rigorously understood via a \emph{reduction} of the $GL(4)$-frame bundle to a $SO(1,3)$-subbundle.
This gives mathematical ground to the heuristic of seeing the reduction of $\Gamma$ to $A$ as a partial gauge fixing of the gauge freedom of coordinate changes. } 

Manifestly, the dressed fields $\{A^u, \phi^u\}$ are \textit{not} a point in the gauge $\K$-orbit $\O^\K_{\{A, \phi\}} \subset \O_{\{A, \phi\}}$ of $\{A, \phi\}$. Thus $\{A^u, \phi^u\}$ cannot be confused with a gauge-fixing of $\{A, \phi\}$, i.e. a point on a gauge-fixing slice $\S$. 
As a matter of mathematical fact, contrary to the action of $\H$ (and any of its subgroups) and therefore contrary to the operation of gauge-fixing, the dressing operation is not a mapping from field space to itself, but a mapping from field space to another mathematical space: the space of dressed fields, denoted $\Phi^u$.
Symbolically we may write
\begin{equation}
\label{GF-map}
\begin{aligned}
 \text{Action of } \H\  : \quad\Phi\  &\rarrow\ \Phi, \\   
 \hookrightarrow\ \  \text{GF}\ : \quad \Phi\  &\rarrow \ \S \subset \Phi, 
 \end{aligned}
 \end{equation}
 \begin{equation}
 \label{Dressing-map}
 \begin{aligned}
 \text{Dressing}\ :\quad \Phi\  &\rarrow\  \Phi^u, \\
 \{A, \phi\} &\mapsto \{A^u, \phi^u\}
\end{aligned}
\end{equation}

Nonetheless, there is  naturally a relation between gauge-fixing and dressing. 
In the case of a complete symmetry reduction via a $\H$-dressing field, 
$\Phi^u$ can indeed be understood as a \emph{coordinatisation} of the moduli space $\M$, or at least of a local region $\U \subset \M$  over which one manages to define a field-dependent dressing field $u: \Phi_{|\U} \rarrow \D r[G, \H]$. 
In such a case, there is indeed a one-to-one mapping $(\Phi_{|\U})^u \leftrightarrow \U \subset \M$, $\{A^u, \phi^u\} \leftrightarrow \{[A], [\phi]\}$ (a coordinate chart).

A gauge-fixing section over that same region, $\s: \U   \rarrow \S \subset \Phi_{|\U}$, establishes a one-to-one mapping $\U \leftrightarrow \S$. 
So there is an isomorphism of spaces $(\Phi_{|\U})^u \simeq \S\subset \Phi_{|\U}$, and yet as mathematical spaces $(\Phi_{|\U})^u \neq \S$. 
DFM thus achieves 
 a comparable goal
 as gauge-fixing (exhibiting physical d.o.f.), but in a technically and conceptually distinct way.\footnote{
In Appendix \ref{AppendixB}, we review the link between DFM and another understanding of gauge-fixing, or gauge choice: the selection of a local trivialisation of the $H$-principal bundle $P$ on which the $\H$-gauge field theory is based. 
This amounts to a choice of bundle coordinates, i.e. fixing the \emph{passive} gauge transformations.
 The main text deals with \emph{active} gauge transformations, coming from the group $\Aut_v(P)$ of vertical automorphisms of $P$ 
 (see Appendix~\ref{AppendixA}).
} 

  This is bound to have interpretive consequences.

\section{Gauge fixing \emph{vs.} dressings in the physics literature}
\label{Gauge fixing vs dressings in the physics literature}

We draw from the physics literature to illustrate these interpretive consequences.

\paragraph{Explicit \emph{vs.} tacit gauge-fixing}
Let us first emphasize a  subtlety concerning the stand 
one adopts towards the gauge-fixing condition. 
As reminded earlier, a gauge-fixing is a local section,  a slice, in field space: 
$\S\defeq \{(A, \phi) \in \Phi_{|\,\U}\, |\, \C(A^\gamma, \phi^\gamma)=0 \}$.
Now, one may either \emph{tacitly assume} the condition to hold, i.e. admit that there are representatives satisfying the condition $\C(A^\gamma, \phi^\gamma)=0$. 
Or one may rather claim that the gauge fixing slice can be reached from any arbitrary chosen point $\{A, \phi\}$ in the orbit $\O$ via some gauge group element, which must then be field-dependent, $\gamma=\gamma[A, \phi]$.\footnote{Such $\Phi$-dependent elements of $\H$ constitute the gauge group of field space as principal bundle.
See e.g. \cite{Francois2021, Francois-et-al2021} for further details.} 

But then it must be stressed that solving explicitly $\C(A^\gamma, \phi^\gamma)=0$ for $\gamma$ may result in actually building a dressing field rather than a field-dependent element of~$\H$. The fact of the matter must be checked by deriving the gauge transformation of the solution: i.e. computing 
\begin{align}
\label{field-dep-GT}
\gamma[A, \phi]^\eta\defeq  \gamma[A^\eta, \phi^\eta], \quad \text{for } \eta \in \H. 
\end{align}
If $\gamma[A, \phi]^\eta= \eta\-\gamma[A, \phi] \eta$, one is  dealing with a genuine field-dependent $\H$-element. 
If $\gamma[A, \phi]^\eta= \eta\-\gamma[A, \phi]$, the explicit solution of the ``gauge-fixing" condition is actually a dressing field, $\gamma[A, \phi]=u[A, \phi]$. 
\medskip

Let us for example consider the 
Lorenz gauge in 
electrodynamics. 
For an EM gauge potential $A_\mu$, transforming via $\gamma=\exp{i\lambda} \in \H=\U(1)$ as 
$A_\mu^\gamma=A_\mu +\d_\mu \lambda$, the Lorenz gauge requires $\C(A)\defeq\d^\mu A_\mu=0$ (the index of $\d_\nu$ being raised via the Minkowski metric $\upeta^{\mu\nu}$). Written as an explicit gauge fixing condition it is 
$\C(A^\gamma)\defeq \d^\mu (A_\mu+\d_\mu \lambda)=0$.

It is one thing to assume the existence of a $A_\mu$ s.t. $\C(A)=0$ holds, and simplify computations by neglecting all  $\d^\mu A_\mu$'s encountered.
But if one explicitly solves $\C(A^\gamma)=0$ for $\gamma$, one finds
\begin{equation}
 \lambda[A]\defeq -\Box\-(\d^\mu A_\mu),   
\end{equation}
where $\Box=\d^\mu\d_\mu$. Checking the gauge transformation, under $\eta=\exp{i \alpha} \in \U(1)$, of this object one easily finds
\begin{align*}
 \lambda[A]^\eta \defeq &\,  \lambda[A^\eta]
  =-\Box\-(\d^\mu A_\mu^\eta)
 = -\Box\-\big (\d^\mu (A_\mu + \d_\mu \alpha)  \big) 
 = -\Box\-(\d^\mu A_\mu) - \alpha \\
  =&\, \lambda[A] -\alpha.
\end{align*}
This implies that $\gamma[A]^\eta=\exp{\{i \lambda[A]^\eta\}}=\eta\-\gamma[A]$. 

Therefore, the solution of the Lorenz ``gauge" condition $\C(A^\gamma)=0$ is  not a field-dependent element of the gauge group $\U(1)$, but~a field-dependent dressing field 
$\gamma[A]=u[A] \in \D r\big[U(1), \U(1)\big]$.
In turn, this implies that 
\begin{align}
 A_\mu^{u[A]}\defeq A_\mu +\d_\mu \lambda[A]   
\end{align}
is not a point in the $\U(1)$-orbit $\O_{\{A\}}$, i.e. not a gauge-fixed $\U(1)$-gauge potential, but  indeed a $\U(1)$-invariant dressed field exemplifying \eqref{dressed-fields}: 
\begin{equation}
\begin{aligned}
&(A_\mu^{u[A]})^\eta = (A_\mu^\eta)^{u[A^\eta]}= (A_\mu^\eta)^{\eta\- u[A]}= (A_\mu^{u[A]}), \\[1mm]
&(A_\mu + \d_\mu \lambda[A])^\eta = (A_\mu + \d_\mu \alpha) + \d_\mu(\lambda[A]-\alpha) = A_\mu + \d_\mu \lambda[A]. 
\end{aligned}
\end{equation}
Notice there are no residual gauge symmetries here (not even harmonic). 

If a spinorial matter field $\phi=\psi$ is considered, with $\U(1)$-gauge transformation $\psi^\eta= \eta\- \psi$, its dressed counterpart is 
\begin{equation}
\psi^{u[A]}\defeq u[A]\- \psi =\exp{\{-i \lambda[A]\}} \psi.
\end{equation}
Again, the latter is not a point in the $\U(1)$-orbit $\O_{\{\psi\}}$ of $\psi$, i.e. it is not a gauge-fixed matter field: 
its $\U(1)$-invariance is easily checked.

So,  $\{A^{u[A]}, \psi^{u[A]}\}$ together are gauge-invariant dressed variables in terms of which one can rewrite the QED Lagrangian: 
\begin{equation}
\label{QED-Lagragangian}
\begin{aligned}
L_{QED}(A_\mu, \psi) &= L_{Maxwell}(A_\mu) +L_{Dirac}(A_\mu, \psi),\\[1.5mm]
    &=\sfrac{-1}{4}\,\,F_{\mu\nu}F^{\mu\nu} 
      + \b\psi (i\slashed D -m)\psi.  
\end{aligned}
\end{equation}
The Dirac operator is built from the covariant derivative $\slashed D  \defeq \upgamma^\mu D_\mu =\upgamma^\mu(\d_\mu + A_\mu )$, where $\upgamma_\mu$ is a Dirac gamma matrix (not a gauge element).  
Using the formal property $L_{QED}(A_\mu^\gamma, \psi^\gamma)=L_{QED}(A_\mu, \psi)$, we get to write:
\begin{equation}
\label{Dressed-QED}
\begin{aligned}
L_{QED}(A_\mu, \psi)&=L_{QED}\big(A_\mu^{u[A]}, \psi^{u[A]}\big) \\[1.5mm]
&= L_{Maxwell}\big(A_\mu^{u[A]}\big) +L_{Dirac}\big(A_\mu^{u[A]}, \psi^{u[A]}\big),\\[1.5mm]
    &=\sfrac{-1}{4}\,\,F_{\mu\nu}^{u[A]}\{F^{u[A]}\}^{\mu\nu} 
      + \b\psi^{u[A]} (i\slashed D^{u[A]} -m)\psi^{u[A]}.  
\end{aligned}
\end{equation}

One may observe that $\{A^{u[A]}, \psi^{u[A]}\}$ are, in a strict field-theoretic sense, \emph{non-local} variables: Since the dressing field $u[A]$ is non-local itself, involving the inversion (integration) of a differential operator -- the Dalembertian -- the dressed variables  assign invariant physical d.o.f. to infinitesimal \emph{regions} of spacetime, not to points.
Invariance has been achieved at the expense of locality: the message seems to be that physical EM properties are fundamentally non-local, i.e. not attributable to points, only to regions. 
So, the dressed theory \eqref{Dressed-QED} appears to be non-local. 

This is consistent with both the holonomy formulation of QED (and gauge theory more generally), see \cite{Healey2009}, and with the message from the Aharonov-Bohm effect, illustrating the non-local, region-like, supervenience of invariant EM properties and effects -- in spinorial QED, see our remark below on scalar QED. 

Yet, notice that since we are in the abelian case:
\begin{equation}
\begin{aligned}
F_{\mu\nu}^{u[A]}=\d_\mu A_\nu^{u[A]} - \d_\nu A_\mu^{u[A]}=\d_\mu A_\nu - \d_\nu A_\mu = F_{\mu\nu},
\end{aligned}
\end{equation}
which is no surprise as the electromagnetic field strength is already $\U(1)$-invariant. 
So, the non-locality in $A^{u[A]}_\mu$  disappears from $F_{\mu\nu}^{u[A]}$, which remains a local invariant field carrying locally propagating  physical d.o.f.
The same happens in the physical matter current 
\begin{equation}
\begin{aligned}
J_\mu^{u[A]}\defeq \b\psi^{u[A]}\upgamma_\mu \psi^{u[A]} =\b\psi \upgamma_\mu \psi = J_\mu.
\end{aligned}
\end{equation}
  Which is to be expected as the latter sources the EM field according to the field equation of the theory: $\d^\mu F_{\mu\nu}^{u[A]} = J_\mu^{u[A]}$.\footnote{Identical to $\d^\mu F_{\mu\nu} = J_\mu$, the Abelian Yang-Mills equation. 
  The sourceless equation $\d_{[\mu} F_{\alpha\beta]}=0$ is just the Abelian Bianchi identity, $dF=0$ in the language of differential forms.  
  It is a geometric data, i.e. kinematically satisfied, no Lagrangian is needed to derive it.}
 From which follows that in vacuum, or far away from the source, one has the relativistic wave equation \mbox{$\Box F_{\mu\nu}^{u[A]}=0$} (identical to the bare version $\Box F_{\mu\nu}=0$).
 The field equation thus implies that, despite the \mbox{fundamental} EM d.o.f. being non-local and assigned to regions of spacetime, they still propagate causally from region to region,  at finite speed along the light cone, in accordance with the spirit and basic desiderata of the \emph{relativistic} field theoretic program. 
 The insight carries over to non-Abelian~gauge~theory.
\medskip

We remark that 
a possible interpretation of $\psi^{u[A]}$ is that of a bare matter (e.g. electron-positron) field intertwined with its EM field. 
This interpretation is  adopted by  \cite{Dirac55, Dirac58} in his attempt at quantizing a gauge-invariant reformulation of QED.  
\cite{Dirac55}~interpreted the quantum field $\h \psi^{u[A]}$ as\footnote{For a particular  solution for $u[A]$ that takes the form of a Coulombic electric field. 
See eq.[16] (Dirac's version of $\psi^{u[A]}$), eq.[17] (the dressing field, a.k.a. Dirac phase), and eq.[19].} 
\begin{quote}
    `` 
    [...] the operator of creation of an electron \emph{together with its Coulomb field}, or possibly the operator of absorption of a positron \emph{together with its Coulomb field}. 
    It is to be contrasted with the operator $\h\psi$ which gives the creation or absorption of a bare particle. 
    \emph{A theory that works entirely with gauge-invariant operators has its electrons and positrons always accompanied by Coulomb fields around them}, which is very reasonable from the physical point of view." (657)
\end{quote}
The emphasis is Dirac's. 
Following him, others have applied the idea in various contexts, invariant field variables often being dubbed ``Dirac variables." 
\cite{McMullan-Lavelle97} e.g. used this strategy to define  physical quarks and gluons states in QCD. We will further comment on their work further down below.

\paragraph{On the trade-off gauge-invariance \emph{vs} locality}
Before following-up on the above point, 
it is perhaps worth observing that one can distinguish (at least) two classes of gauge symmetries  according to whether or not they can be removed without losing the -- field theoretic, not relativistic -- locality of the basic field variables,  hence of the theory itself. 
In \cite{Pitts2008, Pitts2009}, gauge symmetries whose removal don't compromise locality have been called \emph{artificial} (or \emph{fake} in \cite{Jackiw-Pi2015}), while gauge symmetries for which a trade-off between gauge-invariance and locality appears unavoidable are said \emph{substantive} (or \mbox{\emph{substantial} in \cite{Francois2018}).} 

This distinctive criterion arises as one answer to a ``generalised  Kretschmann objection" that can be leveraged against the gauge principle, on the model of the  Kretschmann objection leveraged against the general covariance principle in~GR: 
Given that any field theory can be --  through technical tools such as the Stueckelberg mechanism, see e.g. \cite{Ruegg-Ruiz} and below -- endowed with a gauge symmetry, how could the gauge principle be a constraining heuristics to find empirically adequate theories, and what physical insight would it encode?

A possible answer is that not all gauge symmetries, nor all instances of general covariance, have physical signature. Only the substantive ones do, and one needs only identify their physical signature(s) as distinctive criterion(s). The trade-off gauge-invariance \emph{vs.} non-locality is often accepted as a signature of ``genuine", physically contentful,  gauge symmetries. 
DFM is advocated as a tool to make the criterion operational in \cite{Francois2018}, 
but see \cite{Teh-et-al2021} for a critical appraisal. 

In that context, the $\U(1)$-symmetry of \emph{spinorial} QED can be deemed substantive, since it is usually admitted that eliminating it results in the theory being non-local.
This may be expressed by saying that no (covariant) local dressing field has been found, as the above discussion illustrated. 
Not so for $\U(1)$ in \emph{scalar} electrodynamics. 
Similarly, the $\SU(3)$ gauge symmetry of QCD can be deemed substantive, but $\SU(2)$ in the electroweak sector of the Standard Model may be suspected of being artificial.  
These remarks may be kept in mind as we resume our brief overview of dressings as compared to gauge-fixing in physics.

 \paragraph{Dressings in the gauge theory literature}
The earliest (restrospective) instance of DFM is probably the tetrad in GR, as mentioned above. 
The next might be the Stueckelberg mechanism:  from the definition \eqref{GT-dressing} one may recognise Stueckelberg fields, reviewed e.g. in  \cite{Ruegg-Ruiz}, as (ad hoc, local) dressing fields.
In that form, DFM is found in modern textbook discussions of gauge symmetry: 
 See \cite{Banks2008} section 4.3.1  and   \cite{Schwartz2013} section 8.6. 
 Goldstone bosons, as described e.g. in  \cite{Banks2008} section 7.4,  are also seen to be dressing fields. 
%
\mbox{Relatedly}, dressing fields  feature often 
in massive Yang-Mills theory, see again \cite{Ruegg-Ruiz}, or massive gravity, see \cite{deRham2014}. 

DFM is also relevant  to the gravitational dressings as proposed in \cite{Giddings-Donnelly2016-bis,Giddings2019, Giddings-Weinberg2020}, and to the edge mode literature as initiated in \cite{DonnellyFreidel2016, Freidel-et-al2020-1}. 
Other instances of DFM are easily recognised in the literature on
  gauge anomalies in quantum field theory (QFT) and their BRST treatment, such as \cite{Stora1984} and \cite{Manes-Stora-Zumino1985}, see also \cite{Bertlmann} or \cite{Bonora2023} section 15.1 (on  Wess-Zumino counterterms).
  
In these contexts, conflation of dressings with gauge-fixings, though less tempting, still happens. Which naturally tends to cloud the conceptual analysis of the technical apparatus. 
\medskip

Scalar QED, where matter is described by a complex scalar field $\phi$ with $\U(1)$ transformation $\phi^\gamma =\gamma\- \phi$, is another area where  the DFM applies naturally: 
The polar decomposition of the scalar $\phi=\rho\, e^{i\theta}$ allows to extract the (local) field-dependent dressing field $u[\phi]=e^{i\theta}$, from which one gets the dressed fields $\phi^{u[\phi]}=\rho$ and $A_\mu^{u[\phi]}=A_\mu+ \d_\mu \theta$.  
Within this framework one may model the Aharonov-Bohm effect, see \cite{Wallace2014} and \cite{Francois2018}, or the Abelian Higgs model, see e.g. the textbook by \cite{Rubakov1999} section 6.1, and by extension, some form of supraconductivity, as in \cite{WeinbergVol2} section 21.6. 

The gauge symmetry $\U(1)$ being artificial, $u[\phi]$ being local,
there is no puzzling non-locality in the AB effect (while there still is in the more fundamental spinorial QED), nor is there spontaneous gauge symmetry breaking (SSB) in the Abelian Higgs mechanism and in supraconductivity thus modeled. 
\bigskip

As one might expect, DFM has been shown to be relevant to more general models with SSB, notably in the electroweak (EW) model based on the  group $\U(1) \times \SU(2)$. $\SU(2)$-invariant variables for the model, which are of the form \eqref{dressed-fields}, have been proposed by
\cite{Banks-Rabinovici1979}, and then famously by 
\cite{Frohlich-Morchio-Strocchi80, Frohlich-Morchio-Strocchi81}. 
The latter approach to gauge-invariant perturbative formulation of SSB models and gauge theories, since named FMS, is drawing renewed interest, see e.g. \cite{Maas2019} for a review, also \cite{Maas-Sondenheimer2020} and \cite{Sondenheimer2020}, as well as \cite{Berghofer-et-al2023}
chap.6. 

The treatment of the  EW model by dressing can be found in \cite{Berghofer-et-al2023} chap.5 (see also \cite{Attard_et_al2017}, for focus on the Yang-Mills sector, excluding fermions). 
It dispenses with the notion of SSB altogether: 
The~$\SU(2)$ symmetry is eliminated, and $\SU(2)$-invariant physical fields  
displayed,
in both massless and massive phases of the theory. 
The phase transition of the EW vacuum is shown to be the only physically operative mechanism of mass generation.
This is achieved by extracting a local, field-dependent $\SU(2)$-dressing field $u[\phi]$  from the $\CC^2$-scalar field $\phi$: the $\SU(2)$-gauge symmetry may  thus be  deemed artificial in the model.

The construction of $u[\phi]$ is almost universally misinterpreted as a gauge-fixing, usually known as ``unitary gauge."
The corresponding dressing operation \eqref{dressed-fields} of the gauge potential is as universally understood as Goldstone bosons arising from $\SU(2)$-SSB being ``absorbed" (or, even less seriously, ``eaten") by the gauge potential. 
A  narrative accepted largely uncritically for long, which is puzzling for many reasons. 
One being that seldom does one hear about fermions eating the same goldstone bosons, and yet matter fields are dressed too \eqref{dressed-fields} to get physical fermion fields (leptons and quarks).

Lattice treatments sometimes explicitly exploit a dressed formulation, see e.g. the review by \cite{Creutz2024} 
featuring dressed fermions (eq.16-17). 
This is to be expected, as it is consistent with Elitzur's theorem stating that gauge SSB is impossible in lattice formulations of gauge theory (\cite{Elitzur1975}). A result in manifest tension with the popular understanding of the EW model in terms of SSB.

The EW model thus provides a case study of a long-lived misinterpretation of one's formalism, stemming largely from conflating dressing and gauge-fixing. 
Yet,~it is worth pointing out that treatments of models of ``massive gauge fields" without SSB  were suggested by \cite{Higgs66} and \cite{Kibble67} even before the ground-breaking paper by \cite{Weinberg1967}. \cite{Higgs66} indeed insists that 
``\emph{it~must be possible to rewrite the theory in a form in which only gauge-invariant variables appear}" (compare his eq.(22)-(23) to \eqref{dressed-fields}-\eqref{dressed-Lagrangian} above), while \cite{Kibble67} observes that 
``\emph{It is perfectly possible to describe it} [the model studied] \emph{without ever introducing the notion of symmetry breaking}" (compare eq.(66) to \eqref{dressed-Lagrangian}).

The DFM treatment of the EW model reconnects with these original insights,  fulfills the technical and conceptual expectations laid out twenty years ago by \cite{Earman2004}, and reconciles the continuum formulation  with the lattice approach.

We observe that, taking the Dirac line of interpretation of the dressing in the EW context, one may see $A^{u[\phi]}$ and $\psi^{u[\phi]}$ as respectively the gauge fields and matter fields immersed in the dynamical background provided by the scalar field $\phi$.\footnote{
This is reminiscent of the popular cartoon depiction of the Higgs mechanism meant to illustrate the effective mass acquired by particles through interaction with the scalar field. 
Yet, the suggested interpretation does not directly make that point and only incidentally relates to it.  
}
\medskip

Finally, after QED and the EW theory, we meet instances of the DFM in QCD. 
Let us mention two, both speculative to distinct degrees. 
\smallskip

The first, most speculative, relates to the  ``strong CP violation problem" in QCD, more precisely to one of its  solution proposed by 
\cite{Peccei-Quinn1977}, \cite{Wilczek1978} and \cite{Weinberg1978}: the axion field $a$. 
The latter is a (pseudo-) scalar field, seen as Goldstone-type boson associated to a postulated new global chiral symmetry of the SM, noted $U(1)_{PQ}$. The model is yet another case of SSB: as $U(1)_{PQ}$ is broken, solving the CP problem, the axion acquires a small mass, making a long-lived particle and thus a prime candidate of dark matter. Naturally, attempts are made to derive a gauge version $\U(1)_{PQ}$ of this symmetry.
We won't go into more details, for which see e.g. \cite{DiLuzio-et-al2020} and \cite{Fukuda-et-al2017}, save to say that the defining transformation of the axion $a$ makes it a local $\U(1)_{PQ}$-dressing field.
\smallskip

The second, more solidly  grounded example, is found in the seminal Physics Report by \cite{McMullan-Lavelle97}. 
A work bound to be consulted by anyone interested in gauge-invariant formulations of gauge theories, and of the SM in particular. 
Its main goal is to recover the constituent quarks, accounting for the spectrum of hadron states, from the Lagrangian quarks of QCD. 
The key idea to achieve this is to define constituent quarks and gluons as  gauge-invariant states with well-defined color charge: This in turn is boiled down to finding a ``gluonic dressing," $u[A]$, via which the $\SU(3)$-invariant quark $\psi^{u[A]}$ and gluon $A^{u[A]}$ fields are obtained. 
The dressing $u[A]$ is built  perturbatively, and is highly non-local: it is a non-abelian version of the ``Lorenz condition" analysed above.
Naturally, the authors, inspired by Dirac, embrace the interpretation that the physical quark $\psi^{u[A]}$ is a bare quark shrouded in its gluonic field, and likewise for any valence constituent gluons $A^{u[A]}$. 

We remark that the terminology ``dressing field" is abundantly used throughout.\footnote{\cite{GaugeInvCompFields}, initiating  the DFM, came-up with the same terminology independently.}
The definition of dressing fields in the QCD context, as given by eq.(5.1) and the text above it, clearly states its gauge field dependence as a desideratum.

The authors also put forward the proposal that quark/gluon confinement could be due to Gribov-Singer ambiguities. They argue for it on the basis of a claim that dressings are the same thing as gauge-fixings.
In view of the core thesis of this paper, we must challenge this claim -- Yet, our challenge leaves the proposal essentially intact. 
Analysing how the conflation with dressing arises in this work is instructive as it illustrates a key point that may be overlooked when thinking about gauge-fixing in gauge theory. 
A detailed discussion can be found in \cite{Francois2014} appendix A.1, so we will not repeat it, except to highlight this key point.

In section 5 of their work, Lavelle and McMullan 
specify their goal “to demonstrate that the existence of such a dressing field is equivalent to finding a gauge fixing condition” (p.23). 
Shortly after, we read  
``We have argued that the QCD effective Lagrangian  must contain a gauge fixing term [...] 
generically a function of the potentials, $\chi^a(A)$ [...].
Dressing the quark will involve a further gauge fixing which is \emph{not} 
this Lagrangian gauge fixing, and we will henceforth refer to it as the \emph{dressing gauge fixing}." The emphasis is theirs.  

Let us start by noticing that the coexistence of two distinct ``gauge-fixings" is  curious. If a gauge-fixing condition adequately selects representatives in gauge orbits,  without residual gauge symmetry, why the need for another ``dressing gauge-fixing"? And if full gauge-invariance is achieved via dressing, why the need for a ``Lagrangian gauge-fixing," and how is it even achieved without gauge symmetry left? Regarding the purpose of exhibiting the physical gauge-invariant d.o.f. of a gauge theory, there seems to be a conceptual redundancy in~this~coexistence. 

But setting this aside, how do Lavelle and McMullan purport to establish this notion of ``dressing gauge-fixing"? The structure of the argument is as follows: One considers a functional equation $\C(A^u, \phi^u)=0$ understood to define (if only tacitly) the field-dependent $H$-valued field $u=u[A, \phi]$ -- with $H=SU(3)$ in QCD. 
The~condition $\C=0$ is understood to define a gauge-fixing, i.e. a slice $\S$ in field space $\Phi$, as in 
\eqref{GF} above.
Now the crucial assumption is this:
one \emph{imposes} the invariance of the condition under $\H$-gauge transformations: $\C^\gamma=\C$,
i.e.  
\begin{align}
\label{C-inv}
 \C\big( \{A^\gamma\}^{u^\gamma}, \{\phi^\gamma\}^{u^\gamma}\big)=\C(A^u, \phi^u)\quad  \text{ for } \gamma \in \H. 
\end{align}
Naturally this is equivalent to asking the $\H$-invariance of the field $A^u$ and $\phi^u$, and consequently that the field $u$ gauge transforms as $u^\gamma =\gamma\- u$. It would appear that one thus finds a dressing field $u=u[A, \phi]$ from a gauge-fixing condition $\C=0$.

This contradicts our previous statements. Where is the catch? 
The assumption \eqref{C-inv}, usually made tacitly, is at fault! 
As discussed in section \ref{Basics of classical gauge field theories}, a gauge-fixing condition is the image of a local section $\s : \U \rarrow \Phi_{|\U}$ of field space seen as a $\H$-principal bundle, so that $\S=\{\{A, \phi\} \in \Phi\, |\, \C(A^\gamma, \phi^\gamma)=0\} = \text{Im}(\s)$. 
Now, the fact to remember is that \emph{a local section cannot (generically) be invariant under the action of $\H$} which is (generically) assumed \emph{free}~on~$\Phi$.    
If $\C=0$ is to define a genuine gauge fixing, \eqref{C-inv} cannot hold. Instead, one should have
\begin{align}
\label{C-not-inv}
 \C\big( \{A^\gamma\}^{u^\gamma}, \{\phi^\gamma\}^{u^\gamma}\big)
 =\C\big(\{A^u\}^\gamma, \{\phi^u\}^\gamma\big)
 \quad  \text{ for } \gamma \in \H. 
\end{align}
In other words, the gauge-fixing slice must move \emph{covariantly} (\emph{equivariantly}, in the terminology of differential geometry) as a functional of the fields $\{A, \phi\}$.
Condition \eqref{C-not-inv} is equivalent to requiring $\{A^\gamma\}^{u^\gamma}=\{A^u\}^\gamma$ and $\{\phi^\gamma\}^{u^\gamma}=\{\phi^u\}^\gamma$, which in turn implies $\gamma u^\gamma = u \gamma$, i.e. $u^\gamma =\gamma \- u \gamma$. 
This transformation for $u=u[A, \phi]$ implies that it is a field-dependent element of $\H$ 
(remember the definition \eqref{Gauge-group}), i.e a genuine field-dependent gauge transformation, and it is indeed compatible with  $\S=\{\{A, \phi\} \in \Phi\, |\, \C(A^u, \phi^u)=0\}$ being a gauge-fixing: $A^u$ is indeed a gauge potential, $\H$-transforming as such, and $\phi^u$ $\H$-transforming as $\phi$. 

As we've alluded to in the introduction, this McMullan-Lavelle position, articulated around \eqref{C-inv}, is also argued for e.g. in \cite{Struyve2011} (section 5), \cite{Wallace2024}, and \cite{Gomes2024},  in a slightly different but technically equivalent way: 
The gauge-fixing slice $\S$ defined by $\C=0$ is seen as a target to be projected on, or reachable, from any points $\{A, \phi\}$  of orbits  intersecting it, via a \emph{field-dependent} ``gauge group element" $u[A, \phi]$, which again amounts to requiring \eqref{C-inv} to hold.\footnote{The emphasis here is important as indeed, by assumption, any given point of a gauge orbit is (generically) reachable from any other via some \emph{field-independent} gauge group element. This statement is key to the possibility of establishing a mapping between a gauge orbit and the gauge group -- which is 1:1 if the action of the gauge group on the orbit is free an transitive. 
It is the standard argument in bundle theory by which one can claim that the fibers of a principal bundle are homeomorphic to its structure group,  since the action of the structure group on the fibers is by definition free and transitive.} 
This view cannot hold as we just saw, since   \eqref{C-inv} produces dressing fields and as stressed in section \ref{DFM in a nutshell} around eqs.\eqref{GF-map}-\eqref{Dressing-map}, dressing fields are not elements of the gauge group, $u\notin \H$, so that 
$\{A^u, \phi^u\} \notin \Phi$ (thus $\notin \S$).
To dress is to produce a coordinatization of (a patch $\U$ of) the space of orbits $\M$ merely isomorphic~to~$\S$.
We thus see that this particular conflation between gauge-fixing and dressing arises from overlooking a simple fact about the former. 
Once rectified, their distinct mathematical nature is made clear.

Does this threaten the more ambitious proposition of Lavelle and McMullan that the Gribov-Singer obstruction may be related to confinement?
Actually, the merit of the proposition remains unchanged: it is not undermined by the fact that it is not tied to gauge-fixing. 
Indeed, as observed at the end of section \ref{DFM in a nutshell}, though mathematically different, there is a clear relation between dressings and gauge-fixings in the form of the (abstract)
isomorphism of spaces $(\Phi_{|\U})^u \simeq \S\subset \Phi_{|\U}$.
The Gribov-Singer obstruction characterizes topological constraints of the field space $\Phi$ that apply to the DFM as much as to gauge-fixings. 
It thus remains true that such obstruction may impede the definition of a non-perturbative dressing field $u=u[A]$, and thus  of non-perturbative $\SU(3)$-invariant dressed quark and gluon fields $\psi^{u[A]}$ and~$A^{u[A]}$. 
That this fact may relate to confinement remains to be further argued though. 

Finally, let us consider another remark made by the authors which is an occasion to address another potential source of confusion between dressing and gauge-fixing. It~is claimed
``[...] the dressing gauge is the gauge that removes the explicit dressing from the quarks, resulting in the identification of the Lagrangian fermion with the physical quark field" (p.23-24).
If we  reject this claim  on account of the fact that neither a dressing  nor  the inverse ``undressing" operation are  gauge-fixings, still there is an observation to make:
Abstractly, it is  conceivable that  an element  of the gauge group $\gamma \in \H$ coincides  with a $\H$-dressing field $u \in \D r[H, \H]$, so that $\gamma=u$, \emph{as~functions on $M$}. 
Yet, since they are objects belonging to different mathematical spaces, supporting different actions of $\H$, this equality cannot be  gauge-invariant.\footnote{It is analogous to equating objects of different covariances in GR;  e.g. a (1,2)-tensor and a connection,  ${T^\alpha}_{\mu\nu}={\Gamma^\alpha}_{\mu\nu}$. Such equality would not be covariant under coordinate changes, or $\Diff(M)$. It is not a well-formed mathematical expression. 
Of course, nobody aware of the fact that, despite their analogue index structures, (1,2)-tensors and connections belong to different mathematical spaces would be tempted to write such an equality.
}

Even then, such ``equality without identity" is possible only for field-independent gauge elements and dressings.
Once one considers field-dependent objects, their equality as functional of fields, $\gamma[A, \phi]=u[A, \phi]$, compels that they are the same mathematical object.
Indeed, the action of the gauge group $\H$ on such functionals is, as stressed by \eqref{field-dep-GT},  explicitly controlled by the $\H$-transformations of the variables $\{A, \phi\}$: i.e. 
$\gamma[A, \phi]^\eta \defeq\gamma[A^\eta, \phi^\eta]$ and 
$u[A, \phi]^\eta \defeq u[A^\eta, \phi^\eta]$. If $\gamma$ and $u$ are the same functional of $\{A, \phi\}$, they have the same $\H$-transformation  and are then the same mathematical object; 
either a field-dependent gauge group element, 
or a field-dependent dressing field (or neither). 

{\color{blue}

}

\section{Conclusion} 
\label{Conclusion} 

In this paper, we have clarified the mathematical distinction between gauge-fixing and dressing, illuminated the conceptual relationship between them, and flagged where potential confusions may arise. 
Through this analysis, we have highlighted facts about gauge-fixing that are often overlooked. 
In particular, we have shown that the explicit solution of the ``Lorenz gauge" results not in a gauge-fixing but in a dressing.  
One may suspect that the gauge field literature contains 
several other instances of  ``gauge-fixings" that are actually dressings.

Given that gauge-fixing and DFM carry significantly different physical interpretations, being aware of the above is indispensable for conducting a proper conceptual analysis of the mathematical constructions encountered in the gauge theory literature. We have demonstrated this by discussing illustrative examples.
The aforementioned differences in interpretational implications notably concern (i) the nature of the gauge symmetry (artificial or substantive) of a given model, (ii) the related issue of locality/non-locality of the fundamental d.o.f., (iii) the presence or not of SSB, and most importantly, (iv) how to identify and interpret the physical, gauge-invariant d.o.f. of gauge theories. 

For anyone interested in gauge-invariant formulations of gauge field theories, (iv) is particularly significant. DFM, intended to streamline such formulations, lends itself to interpretation quite differently than usual gauge-fixing procedures. 
We have encountered such a natural interpretation, with Dirac first in QED, and Lavelle and McMullan  then in QCD, but also in the EW model: 
gauge-invariant fields are ``composite," a bare gauge-variant field enveloped into another.\footnote{Another, not unrelated, interpretation of the formalism of the DFM is explicitly \emph{relational}. Bare, gauge-variant fields are seen as partial observables, in the terminology of \cite{Rovelli2002}, while dressed fields are complete observables (see e.g. \cite{Tamborino2012}, \cite{Rovelli2014}, \cite{Gomes2019} for discussions of relationality in gravity and gauge theory). 
The DFM as presented and discussed here applies to gauge theories with internal gauge groups. It is extended to $\Diff(M)$ in \cite{Francois2023-a}, so as to apply to general relativistic theories. 
The relational interpretation of the framework is both natural and  appealing. 
Of course, such a physical picture is foreign to gauge-fixing.} 
This applies whether it is a physical electron as a bare electron shrouded in its electromagnetic field, or physical quarks and gluons as their bare selves similarly dressed by their gluonic clouds. 
Crucially, gauge-fixing does \textit{not} yield such a physical picture. 


\section*{Acknowledgment}

This research was supported by the Austrian Science Fund (FWF), [P 36542].
JF also acknowledges support from 
OP J.A.C MSCA grant, co-funded by the Czech government Ministry of Education, Youth \& Sports and the EU (number CZ.02.01.01/00/22\_010/0003229).

\appendix
\section{A primer on bundle theory}
\label{AppendixA}

For the reader interested in knowing why the kinematical structure of gauge field theory is the way it is (as given notably by \eqref{Gauge-group}-\eqref{GTgauge-fields}-\eqref{GTgauge-fields2}), we here provide a concise summary of the fundamentals of bundle geometry. 
A number of good references can fill the ommitted  details, we suggest \cite{Hamilton2018,Bertlmann, Sharpe, DeAzc-Izq}. We may observe that bundle geometry is the natural framework for a rigorous formulation of the DFM -- see e.g. \cite{Berghofer-et-al2023, Francois2021, Attard-Francois2016_I}, and \cite{Francois2023-a} for its extension to diffeomorphism symmetry.
\medskip

The fundamental space considered is a principal fiber bundle $P$ over $M$ with structure group $H$, $P\xrightarrow{H}M$. The right action of $H$ on $P$ is free, and foliates it into fibers, the space of fibers $P/H$ is isomorphic to $M$. There is then the canonical projection $\pi:P \rarrow M$, $p \mapsto \pi(p)=x$. 
Points of $M$ are thus seen as having an internal structure, the fibers, so $P$ can be understood as an enriched spacetime.  
Its natural maximal automorphism group is $\Aut(P)$, those diffeomorphisms of $P$ that maps fibers to fibers, thus induces diffeomorphisms of~$M$. 
The subgroup $\Aut_v(P)$ of \emph{vertical} automorphisms, inducing the identity transformation on $M$, is isomorphic to the gauge group $\H(P)$ of $P$:  maps  $\upgamma:P \rarrow H$, $p \mapsto \upgamma(p)$,  s.t. $\upgamma(ph)=h\- \upgamma(p)h$. 
There is a one-to-one association between a vertical automorphism $\psi\in Aut_v(P)$ and a generating element $\upgamma \in \H(P)$.
Given any form $\alpha \in \Omega^\bullet(P)$, its gauge transformation is defined as the pullback action by $\Aut_v(P)$: $\psi^*\alpha$. If the result can be expressed in terms of the generating element $\upgamma \in \H(P)$, one writes: $\alpha^\upgamma \defeq \psi^* \alpha$. In particular, for any $\upeta \in \H(P)$, one has $\upeta^\upgamma =\psi^*\upeta = \upgamma\-\upeta \upgamma$. 

A connection on $P$ is  $\omega \in \Omega^1(P, \LieH)$ with two defining properties\footnote{To state them without further explanation: a connection is $\ad(H)$-equivariant and is a projector on vector fields tangent to the fibers. } implying that $\Aut_v(P)\simeq \H(P)$ acts on it as $\omega^\upgamma \defeq \psi^* \omega =\upgamma\- \omega\upgamma + \upgamma\-d\upgamma$. Its curvature 2-form is $\Omega = d\omega +\sfrac{1}{2}[\omega, \omega] \in \Omega^2(P, \LieH)$, and accordingly transforms as $\Omega^\upgamma =\psi^*\Omega =\upgamma\- \Omega\upgamma$.
Tensorial forms $\beta \in \Omega_\text{tens}^\bullet(P, V)$, are representation-valued forms that are $\rho(H)$-equivariant ($\rho$ a representation morphism for $H$) and horizontal (they vanish when evaluated on vector fields tangent to fibers). This implies that they gauge transform as: $\beta^\upgamma =\psi^*\beta = \rho(\upgamma)\- \beta$. 
The exterior derivative does not preserve this space, $d\beta \notin \Omega_\text{tens}^\bullet(P, V)$. But the covariant derivative $D=d +\rho_*(\omega)$ is designed to: $D\beta \in \Omega_\text{tens}^\bullet(P, V)$, so that $(D\beta)^\upgamma =\psi^* (D\beta)= \rho(\upgamma)\- D\beta$. The curvature is a tensorial 2-form for the $Ad$-representation, it satisfies Bianchi identity $D\Omega=d\Omega +[\omega, \Omega]=0$.  Matter fields are tensorial 0-forms on $P$, $\beta=\upphi \in \Omega_\text{tens}^0(P, V)$, so that $D\beta=D\upphi \in \Omega_\text{tens}^1(P, V)$. Hence their gauge transformations.
\bigskip

A principal bundle is locally trivial, i.e. over $U\subset M$ it is the case that $P_{|U} \simeq U \times H$. 
The map realising the isomorphism is called a local trivialisation and provides bundle coordinates $p \mapsto (x, h)$. 
Over $U$, it is always possible to choose a local section $\sigma : U \rarrow P_{|U}$, $x\mapsto \s(x)\simeq (x, \id_H)$, which then supplies a ``reference" for such bundle coordinates (notably concerning the fiber coordinate). 
Via such a local section, one may pullback forms living on $P_{|U}$ down to $U$. In particular, the fields of a gauge theory are local representatives of global objects on $P$, obtained~by 
\begin{align}
\label{pullback}
A \defeq \s^* \omega \in \Omega^1(U, \LieH), \qquad \phi \defeq \s^* \upphi \in \Omega^0(U, V), \qquad F \defeq \s^* \Omega \in \Omega^2(U, \LieH).
\end{align}
Viewed differently, one may say that via $\s$ one gets a particular bundle-coordinate view from $U\subset M$ of the globally defined objects.
Similarly, one gets the local gauge group on $U$  via $\H = \s^* \H(P)$, giving \eqref{Gauge-group}, while the (local) gauge transformations \eqref{GTgauge-fields}-\eqref{GTgauge-fields2} are obtained via $A^\gamma = \s^* \omega^\upgamma$, $\phi^\gamma = \s^* \upphi^\upgamma$, and $F^\gamma = \s^* \omega^\upgamma$. 
These are \emph{active} local gauge transformations: they are the local view, in a given bundle-coordinate system, of the active $\Aut_v(P)\simeq \H(P)$ transformation relating \emph{different} global objects on $P$. 

\emph{Passive} gauge transformations  arise from changing the local section by which the pullback \eqref{pullback} is operated: by switching $\s \mapsto \s'=\s g$ with $g:U \rarrow H$ a so-called \emph{transition function} of $P$, one gets new local representatives of global objects on $P_{|U}$ related to hold ones by:
\begin{align}
  \label{localGT}
 A':=g^{-1} A g + g^{-1} d g \qquad \text{and} \qquad 
 \phi':=g^{-1}\phi. 
\end{align}
The above are often called ``gluing" relations, under the assumption that $\s$ and $\s'$ are local sections over different yet overlapping regions $U$ and $U'$ s.t. $U\cap U' \neq \emptyset$. 
Each defining a local bundle coordinate system over $P_{|U}$ and $P_{|U'}$ respectively, the function $g:U\cap U' \rarrow H$ realises the junction of them on the overlap $P_{|U} \cap P_{|U'}$ -- i.e. $h'=hg$ for the fiber coordinate -- hence the name ``transition functions" for $g$ maps: they play exactly the same role as transition functions between coordinate charts in standard manifold theory.
Equations \eqref{localGT} are to be understood as relating different views, from distinct bundle-coordinate systems, of the \emph{same} global objects. 
\medskip

Physically, passive gauge transformations are indistinguishable from  active gauge transformations \eqref{GTgauge-fields}, which a priori has profound interpretive consequences.

If the 
gauge principle is understood as requiring invariance of the Lagrangian $L$ (or the action $S$) under passive gauge transformations \eqref{localGT}, it amounts then to asking that it is well defined across $M$ as it glues trivially over overlapping regions $U\cap U'$. 
This in turn implies that it descends from (it is the local representative of) a global object on $P$ that is said ``basic": it is invariant under $\Aut_v(P)\simeq \H(P)$. 
So, asking for passive gauge invariance (of the local Lagrangian/action) automatically implies obtaining active gauge invariance (of its global counterpart), under $\Aut_v(P)$. 
The conceptual meaning of each is as different as invariance under general coordinate changes and under diffeomorphisms $\Diff(M)$ in GR.

\section{Dressings and bundle-coordinates}
\label{AppendixB}

Above, we have briefly evoked the fact that a fiber bundle $P$ is locally trivial over $U\subset M$, so that $P_{|U}\simeq U \times H$. 
The  map $\vphi: P_{|U}\rarrow U \times H$ realising the isomorphism is called a \emph{local trivialisation}, and is said to provide \emph{bundle coordinates}. 
Since both $P$ and $U\times H$ are right $H$-spaces, i.e. they both support a right action of $H$, the local trivialisation must be $H$-equivariant: $\vphi(ph)=\vphi(p)h$, for $h\in H$.  
Explicitly, 
\begin{equation}
\label{loc-triv}
\begin{aligned}
\vphi\ :\  P_{|U} \ &\rarrow\  U \times H, \\
       p \ &\mapsto\  \vphi(p)
              =\big(\pi(p), \b \u(p) \big) 
              = \big(x, \b \u(p) \big), \\
       ph \ &\mapsto\  \vphi(ph)
              =\big(\pi(ph), \b \u(ph) \big)
              =\big(x, \b \u(p) h \big)
              =\big(x, \b \u(p)\big)h.
\end{aligned}
\end{equation}

The $H$-equivariant map $\b \u: P_{|U} \rarrow H$, $p \mapsto \b \u(p)$, s.t. $\b \u(ph)=\b \u(p)h$, is sometimes itself called the local trivialisation of $P_{|U}$ (since it contains all the information of $\vphi$, given that the projection $\pi: P \rarrow M$ is canonical). It can be understood as a $H$-valued tensorial 0-form on $P_{|u}$, and by the discussion in the previous appendix, this means that it transforms under $\H(P)$ as $\b \u^\upgamma =\b \u \upgamma$. Its local representative via a local section $\s: u \rarrow P_{|U}$ is $\b u\defeq \s^* \b \u$, and transforms under the local gauge group $\H$ as $\b u^\gamma =\b u \gamma$. 

Comparing the latter with the definition \eqref{GT-dressing},
we see that $u\defeq \b u\-$, which transforms as $u^\gamma=\gamma\- u$, is a $H$-valued $\H$-dressing field: $u\defeq \b u\- \in \D r[H, \H]$. 
Therefore,  any local trivialisation supplies  a special case of dressing field. 
Yet, writing local representatives via $\s^*$ is still  distinct from dressing these representatives, which are invariant under change of bundle coordinates (i.e. local sections). The reader may consult the Appendix of \cite{Francois2018} for further details on this. 


{
\normalsize 
 \bibliography{biblio-draft}
}

\end{document}